\input{aipcheck}
\documentclass[
    ,final            % use final for the camera ready runs
  ]
  {aipproc}
\layoutstyle{6x9}

\usepackage{graphicx}

\def\bfm#1{\hat{#1}}

\newcommand{\ga}{\gamma}

\newcommand{\De}{\Delta}

\newcommand{\eps}{\epsilon}

\newcommand{\la}{\lambda}
\newcommand{\Lam}{\Lambda}

\newcommand{\p}{\partial}
\newcommand{\lt}{\langle} 
\newcommand{\gt}{\rangle}  % LaTeX: \> already defined
\newcommand{\txt}{\textstyle}
\newcommand{\dsp}{\displaystyle}

\newcommand{\Eqn}[1]{Eq.~(\ref{#1})}  % includes ``Eq.'' in front
\newcommand{\beq}{\begin{equation}}
\newcommand{\eeq}{\end{equation}}
\newcommand{\ba}{\begin{array}}
\newcommand{\ea}{\end{array}}
\newcommand{\bea}{\begin{eqnarray}}
\newcommand{\eea}{\end{eqnarray}}
\newcommand{\bal}{\begin{align}}  % align in amsmath is better than eqnarray
\newcommand{\eal}{\end{align}}
\newcommand{\bi}{\begin{itemize}}  %\setlength{\itemsep}{0\parsep}}
\newcommand{\ei}{\end{itemize}}
\newcommand{\ben}{\begin{enumerate}}  %\setlength{\itemsep}{0\parsep}}
\newcommand{\een}{\end{enumerate}}

\newcommand\hide[1]{}

\newcommand{\ie}{{i.e.}}

\newcommand{\feyn}[1]{
  \setbox0=\hbox{\ensuremath{#1}}
  \hbox to\wd0{\hbox to0pt{\hbox to\wd0{\hss/\hss}\hss}\box0}}
 
\newcommand{\MeV}{\,{\rm MeV}} 
 
\newcommand{\diag}{{\rm diag.}}

\newcommand{\mue}{\mu_{e}}
\newcommand{\ms}{m_s}
\newcommand{\Ms}{M_s}

\begin{document}

\title{Thermal unpairing transitions affected by\\ 
neutrality constraints and chiral dynamics}


\author{Hiroaki Abuki}{
  address={Yukawa Institute for Theoretical Physics, 
  Kyoto University, Kyoto 606-8502, Japan}
}

\author{Teiji Kunihiro}{
  address={Yukawa Institute for Theoretical Physics, 
  Kyoto University, Kyoto 606-8502, Japan}
}

\begin{abstract}
 We discuss the phase structure of homogeneous quark matter under the
 charge neutrality constraints, and present a unified picture of the
 thermal unpairing phase transitions for a wide range of the quark density.
We supplement our discussions by developing the Ginzburg-Landau analysis.
\end{abstract}

\maketitle

Surprisingly rich phases of quark matter under high pressure are being
revealed by extensive efforts
\citep{Shovkovy:2003uu,Alford:2003fq,Abuki:2004zk,%
Abuki:2005ms,Ruster:2005jc,Blaschke:2005uj}.
The key ingredients which are important at moderate density
can be divided into two categories;
(i) the dynamical effects and (ii) the kinematical effects. 
The dynamical effects include  strong coupling effects
\citep{Abuki:2001be,Nishida:2005ds,Kitazawa:2005ux,He:2005bd}, an interplay between the
pairing and chiral dynamics
\citep{Abuki:2004zk,Abuki:2005ms,Ruster:2005jc,Blaschke:2005uj} and 
so on.
In this talk, we focus on the competition of the pairing with the chiral
dynamics and the kinematical effects 
such as (1) the strange quark mass and (2) the charge neutrality
constraints under the $\beta$-equilibrium. 
The kinematical effects will make a stress on the pairing and bring
about the exotic phase called ``{\em gapless}'' superconductivity
at moderate density \citep{Shovkovy:2003uu,Alford:2003fq}; 
the gapless CFL (gCFL) \citep{Alford:2003fq} is one of such examples.
The neutrality constraints are also known to lead to
an interesting complication in the phase diagram
even at finite temperature: For instance, the Ginzburg-Landau analysis
\citep{Iida:2003cc}
shows that the down-quark pairing  phase (dSC phase) 
 consisting of only the  $u$-$d$ and $d$-$s$
pairings may become the second coldest phase at high density. However,
this conclusion seems to be model-dependent.
In fact, the NJL analyses which incorporate the chiral dynamics
\citep{Ruster:2005jc,Blaschke:2005uj} show that the second densest phase
is the up-quark pairing
(uSC) phase consisting of only the  $u$-$d$ and $u$-$s$ pairing.
In this talk, we report our recent work \citep{Abuki:2005ms} which
gives a systematic and unified picture of the thermal unpairing
transitions under the charge neutrality constraints for a wide range of
the quark density.

\begin{figure}[t]
\includegraphics[scale=0.22, clip]{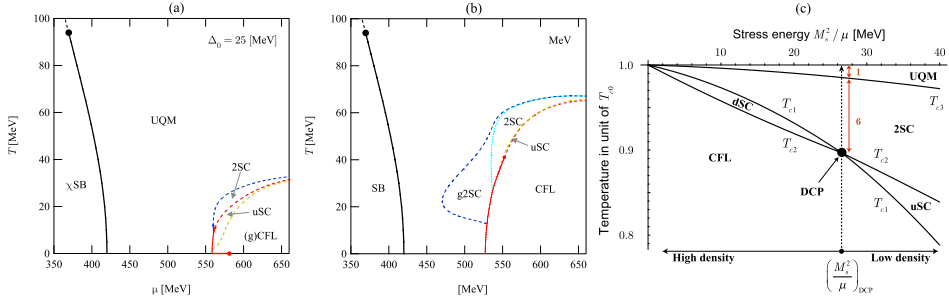}
\caption{Phase diagram calculated with the extremely weak
  $(qq)$ coupling (a) and that with weak coupling (b). In (a), there is
  a small region for the realization of gCFL phase at $T=0$ (bold red
  line). (c) shows $T_{c\eta}$ evaluated by the Ginzburg-Landau
  analysis with the parameter set $G_D/G_S=0.42$ and
  $\mu=500{\,{\rm MeV}}$.
  When the coupling strength is increased, $(M_s^2/\mu)_{\rm DCP}$,
  the intersection of $T_{c2}$ and $T_{c1}$ moves to higher value,
  while the ratio of $T_{c0}-T_{c3}:T_{c0}-T_{\rm DCP}=1:7$ would not be
  affected.\vspace*{-0.55cm}
\label{fig0}
}
\end{figure}

We start with the Lagrangian density ${\mathcal L}=\bar{q}(i\feyn{\p} -
\bfm{m}_0 + \bfm{\mu}\ga_0)q+{\mathcal L}_{\rm int}$ with ${\mathcal
L}_{\rm int}$ being the following 4-fermion coupling \citep{Abuki:2004zk}
\beq
 \txt{\mathcal L}_{\rm int}=G_D\sum_{\eta=1}^{3}%
       \big[(\bar{q}P_\eta^t\bar{q})(^tq\bar{P}_\eta q)\big]%
       + G_S\sum_{\alpha=0}^8\big[(\bar{q}\la_{F\alpha}q)^2 +
       (\bar{q}i\ga_5\la_{F\alpha} q)^2\big].\label{eq:lag}
\eeq
The first term simulates the attractive interaction in
  the color (flavor) anti-triplet and $J^P=0^+$ channel in
  QCD, \ie, $(P_\eta)^{ab}_{ij}=i\ga_5C\eps^{\eta ab}\eps_{\eta ij}$.
See \cite{Abuki:2005ms}, for the other details of notation.
We take the chiral SU(2) limit $\bfm{m}_0=\diag\{0,0,\ms\}$
with $m_s=80\MeV$ fixed.
$\bfm{\mu}$ in the Lagrangian contains the charge chemical potentials
$(\mu_e,\mu_3,\mu_8)$ which couple to the electric and two diagonal
color charge densities as
\beq
 \bfm{\mu}_{ij}^{ab}= \dsp \mu - \mue Q_{ij} + \mu_3 T_3^{ab} + \mu_8
                 T_8^{ab} + \mbox{(off-diagonal part)},
\label{chemicalpot}
\eeq
where we have defined $Q$, $T_3$ and $T_8$ as usual \cite{Abuki:2005ms}.
It can be shown that the chemical potentials for
off-diagonal color densities are unnecessary for the standard diagonal
ansatz for the diquark condensate, \ie, $\langle
q^a_iq^b_j\rangle\sim(P_\eta)^{ab}_{ij}$ with $\eta=1,2,3$
\citep{Abuki:2005ms}.
We determine the phases in $(\mu,T)$-plane by calculating
the effective potential through the mean field approximation with
the condensate fields, $\De_\eta=\frac{G_d}{8}\lt ^tqP_\eta q\gt$, and 
$\bfm{M}-\bfm{m}_0=-\frac{G_s}{2N_c}\diag(\lt\bar{u}u\gt,\lt\bar{d}d\gt,\lt\bar{s}s\gt)$.
We perform the calculation with several values of $G_D/G_S$ with
$G_S$ set so as to reproduce the dynamical quark mass $M=400\MeV$ in the
chiral limit at $\mu=T=0$ for a cutoff $\Lam=800\MeV$; we have
$G_S\Lam^2=2.17$.
Fig.~\ref{fig0}(a) and (b) show the phase diagram for $G_D/G_S\cong0.42$
({\em the extremely weak coupling case}) and that for $G_D/G_S\cong0.63$
({\em the weak coupling case}).
The $\chi$SB denotes the chiral-symmetry broken phase
and the UQM is an abbreviation of the ``{\em unpaired quark matter}''.
For other phases, see \cite{Abuki:2005ms}.
Each of the superconducting phases, CFL, 2SC and uSC, has its gapless
version, gCFL, g2SC, and guSC, where some quasi-quarks become gapless in
the presence of the finite background charge chemical potentials.
As the value of the diquark coupling is increased, these ``{\em
premature}'' gapless phases tend to disappear, and the fully gapped
phases dominate the phase diagram. In fact, we can see that the gCFL
phase at $T=0$ in Fig.~\ref{fig0}(a) is taken over by the UQM phase in
(b), which can be interpreted as a consequence of the competition
between the dynamical and kinematical effects \cite{Abuki:2005ms}.

The reason why the uSC shows up in the phase diagram
instead of the dSC can be nicely understood by the Ginzburg-Landau (GL)
analysis.
We can expand the effective potential in terms of the gap parameters
near the critical temperature $T_{c0}$ which denotes the CFL $\to$ UQM
transition temperature in the symmetric matter with $m_s=0$:
\beq
\txt {\mathcal L}_{\rm GL}=4N[\mu]\big\{-f_i(\Ms)\Delta_i^2%
 +\frac{1}{2}g_{ij}(\Ms)\De_i^2\De_j^2+\cdots\big\},\label{eq:gl}
\eeq
where $N[\mu]=\mu^2/2\pi^2$ is the density of state. 
It is important that under the neutrality constraints, the GL
coefficients become  functions of $\Ms$,  which may be expanded as
\beq
  f_i(\Ms,T)=a_{0i}+a_{2i}\Ms^2+a_{4i}\Ms^4+\cdots,\quad%
  g_{ij}(\Ms,T)=\beta_{0ij}+\beta_{2ij}\Ms^2+\cdots.
\eeq
We can calculate all the coefficients using the Feynman diagrams.
The ultraviolet divergence appears only in $a_{0i}$
as a consequence of the singularity in the diquark propagator
at $T=T_{c0}$ for $\Ms=0$.
We can regularize it by subtracting the equation expressing the Thouless
condition which serves as the mass counter term that guarantees the
second order transition at $T=T_{c0}$.
We can show $a_{0i}=-t$ $(i=1,2,3)$ with $t$ being the reduced
temperature $\frac{T-T_{c0}}{T_{c0}}$, and $\beta_{0ij}=\delta_{ij}%
\frac{7\zeta(3)}{16\pi^2 T_{c0}^2}$.
$a_{2i}$ was derived in \citep{Iida:2003cc}, and $a_{4i}$, 
$(\beta_{2ij})$ was obtained\footnote{One has to take care that there
is a feedback to $\beta_{2ij}$ from the Fermi gas part of thermodynamic
potential.} in \citep{Abuki:2005ms}.
As we show below, ${a_{2i}}$ $(a_{4i},\beta_{2ij})$ causes
 a split of the order of $\Ms^2$ ($\Ms^4$) in the melting temperature;
$T_{c0}\to(T_{c1},T_{c2},T_{c3})$ where $\De_\eta$ vanishes at
$T_{c\eta}$.
When $T$ is increased, 
the first CFL-to-{\em non}-CFL transition with
$\De_{\eta_1}\to0$ takes place at $T=T_{c\eta_1}$. 
We can determine $T_{c\eta_1}$ by solving $T_{c{\eta_1}}={\rm
min}.\{T_{1},T_{2},T_{3}\}$ with $T_{\eta}$ defined by the root of
$\De_\eta^2(T_{\eta})=0$ where $\De_\eta^2(T)=g^{-1}_{\eta j}f_j(\Ms,T)$
is the solution of $\frac{\partial\mathcal L_{\rm
GL}}{\partial\vec{\De}}\Big|_{{}_T}=\vec{0}$. We have
%\vspace*{-0.15cm}
\beq
\ba{rcl}
  \frac{T_\eta}{T_{c0}}&=&1+a_{2\eta}\Ms^2+\Big(a_{4\eta}%
  +\frac{7\zeta(3)}{16\pi^2 T_{c0}^2}\sum_j\beta^{-1}_{2\eta
  j}(a_{2\eta}-a_{2j})\Big)\Ms^4+\cdots,%
\ea
%\vspace*{-0.1cm}
\eeq
by which $T_{c{\eta_1}}$ as a function of $\Ms$ can be determined.
To find the next melting order parameter $\De_{\eta_2}$,
we put $\De_{\eta_1}=0$ into \Eqn{eq:gl} and repeat the same procedure
in two order parameter space.
Finally, we obtain the order of hierarchical melting transitions
($T_{c\eta_1}<T_{c\eta_2}<T_{c\eta_3}$).
We confirmed that $\De_3$ survives at highest temperature 
so that $\eta_3=3$ irrespective of the value of $\Ms$. In contrast,
which of $\De_1$ and $\De_2$ first vanishes with increasing 
$T$ depends on $\Ms$; in fact, we have found the {\em doubly critical}
strange quark mass $\Ms^{\rm DCP}$ above (below) which the uSC (dSC) is
realized as the second coldest phase. 
In Fig.~\ref{fig0}(c), we have given the phase diagram in the
$(\Ms^2/\mu,T)$-plane, obtained by the GL analysis for
$G_D/G_S\cong0.42$ and $\mu=500\MeV$.

In conclusion, we have made an extensive analysis of the  phase
diagram of the quark matter 
and given a unified view on the thermal unpairing transitions.
By extending the earlier work \citep{Iida:2003cc} with the higher
order effects of the strange quark mass on the pairing taken into
account, we have shown how the window for the dSC-realizaion in the 
high density regime tends to close towards lower density.
It should also be stressed that an analytic expression for the doubly
critical point \citep{Fukushima:2004zq} 
can be derived in our framework.

%\endinput
\end{document}